# Room temperature surface piezoelectricity in SrTiO$_3$ ceramics via piezoresponse force microscopy

Andrei Kholkin, Igor Bdikin, Tetyana Ostapchuk*, and Jan Petzelt*

Dept. of Ceramics and Glass Engineering, CICECO, University of Aveiro, 3810-193 Aveiro, Portugal
*Institute of Physics, Czech Academy of Sciences, Prague, Czech Republic

Abstract:

SrTiO$_3$ ceramics are investigated by piezoresponse force microscopy. Piezoelectric contrast is observed on polished surfaces in both vertical and lateral regimes and depends on the grain orientation varying in both sign (polarization direction) and amplitude. The observed contrast is attested to the surface piezoelectricity due to flexoelectric effect (strain gradient-induced polarization) caused by the surface relaxation. The estimated flexoelectric coefficient is approximately one order of magnitude smaller as compared to those recently measured in SrTiO$_3$ single crystals. The observed enhancement of piezoresponse signal at the grain boundaries is explained by the dipole moments associated with inhomogeneous distribution of oxygen vacancies.

SrTiO$_3$ (STO) is one of the most investigated materials of the ferroelectric perovskite-type oxides family due to its large and tunable dielectric permittivities, thermal stability and a variety of physical phenomena ranging from incipient ferroelectricity to superconductivity and giant piezoelectricity at low temperature [1]. At room temperature bulk STO has a cubic (*m3m*) centrosymmetric perovskite structure and does not exhibit bulk piezoelectric effect. It undergoes a nonpolar antiferrodistortive phase transformation to a tetragonal structure with point group 4/*mmm* at ~105 K and exhibits indications of a frustrated ferroelectric transition at ~20 K which however never completes [2]. In addition, the properties of STO are prone to size effects and mechanical stress as was proved by the comparison of ceramics and thin films and single crystals of the same stoichiometry. In particular, ferroelectric state can be induced in STO films by applying biaxial strain and by doping [3, 4]. Macroscopic polarization could also appear in STO crystals subjected to inhomogeneous deformation (flexoelectricity) [5] or temperature (thermopolarization effect [6, 7]). Flexoelectricity can significantly affect the functional properties of high-permittivity materials, especially in complex geometries and as a part of composites [8, 9]. In addition to the possible surface ferroelectricity (due to flexoelectric effect or to oxygen rearrangement [10]) it was shown that ferroelectric polarization may form near the grain boundaries [11] due to segregation of Ti or deficiency in oxygen [12]. Petzelt et al [11, 13, 14, 15] compared the Raman and infrared (IR) reflectivity in STO ceramics of different grain sizes and explained the results by the existence of frozen polarization $P^f$ independent of temperature. Local techniques are obviously necessary to study the polar features related to the STO grain structure and possible dipole moments at the surface. In this work, we used piezoresponse force microscopy to study the polar phenomena and associated piezoelectric effects in pure STO ceramics.

A commercial SFM (PicoPlus$^{TM}$, Agilent Technologies) was used for the local ferroelectric domain studies. The microscope was equipped with an external lock-in amplifier (SR-830, Stanford Research) and a function generator (FG120, Yokogawa), which were used to apply the ac and dc voltages to the surface for poling and imaging acquisition [16]. The amplitude and frequency of the ac voltage were 1–30 V and 50 kHz, respectively. Conducting Si cantilevers (42 Nm$^{-1}$, PPP-NCHR and 0.2 Nm$^{-1}$, PPP-CONTR, Nanosensors) were used for the measurements performed in ambient environment. The tip has the shape of a polygon-based pyramid with the height of 10-15 μm and the effective radius $r_{tip}$ ≈10 nm.

SrTiO$_3$ ceramic samples were prepared using mixed oxide route with raw oxides SrCO$_3$ (Merck, Germany) and TiO$_2$ (Koygo, Japan). Final sintering was done at 1380 ºC for 7 hours and the mean grain size was in the range 1-2 μm. The impurity content was below 120 mol ppm as verified by inductive coupled plasma induced atomic emission spectroscopy. The details of the sintering procedure can be found elsewhere [11].

Figure 1a shows a representative topography image of the polished STO ceramic surface which was smooth enough for the PFM measurements. RMS roughness estimated from the image is around 5-10 nm. Polishing scratches with depths down to several hundred nm and some grains characterized by notable height differences are visible on the image. From both topography and piezoresponse images (Figs.1b and c) we could easily distinguish STO grains with the lateral size expected from SEM. Comparison of the topography and PFM images (Fig. 1) attests that the polishing scratches and other corrugations of the sample surface are not reflected on the corresponding PFM images thus giving an evidence of the piezoelectric nature of the contrast. The observed PFM signal is roughly uniform within the grains and but differs from grain to grain by magnitude (value of the corresponding effective piezoelectric coefficient) and by polarization direction (phase of the signal). These images are almost identical to typical PFM contrast in polycrystalline materials [17] with single domain grains, being much smaller in amplitude. The effective vertical piezoelectric coefficient never exceeded 1-2 pm/V as judged by comparison with commercial thick PZT films prepared by sol-gel [18]. For out-of-plane (OPP) component (Fig. 1b) the contrast is roughly proportional to the effective $d_{33}$ coefficient and determined by the projection of the polarization vector **P** on the normal **N** to the ceramic surface. The dark and bright contrasts correspond to the polarization head directed to the STO bulk or to the surface, respectively. This polarization component apparently varies in both sign (polarization direction) and magnitude (amplitude of the effective piezoresponse) depending on crystallographic orientation. In-plane (IPP) piezoresponse image (Fig. 1c) is complementary to the OPP picture and reflects the effective shear $d_{15}$ piezoelectric coefficient (proportional to the corresponding in-plane component of the polarization). The linearity of the observed contrast with applied ac voltage was verified down to the resolution limit of PFM and also confirmed the piezoelectric nature of the contrast.

The histogram of the OPP PFM signal from image shown on Fig. 1b is represented in Fig. 2a. It has a clear asymmetric shape with a maximum slightly shifted towards the negative value. It means that the number of grains with the polarization head terminated at the free surface exceeds the number of oppositely oriented dipoles. Under large magnification, a shoulder on the positive slope was observed that may represent a maximum in the number of grains having the same effective piezoelectric coefficient. Figure 2b shows a representative cross-section of the IPP piezoresponse taken across three grains depicted in Fig. 1c. Clearly, the contrast is amplified near the grain boundary. This amplification is reflected in both OPP and IPP components and is uncoupled from the surface topography. It is also found that in many cases the increase of the IPP signal is more pronounced that the OPP response. Thus, the grain boundary plays an important role in the piezoelectric contrast observed on the surface of the undoped STO ceramics. These features are in line with Raman and IR measurements revealing forbidden modes characteristic of the polar state of $SrTiO_3$ ceramics [11] interpreted within a modified dead layer model [14]. Thus our measurements give a clear evidence of the polarization existence not only at the STO surface (as judged by the contrast inside the grain) but also its magnification at the grain boundaries. We tried to polarize the surface of the STO ceramics by applying a sufficiently high voltage to the tip (≈100 V) for a long time during scanning (scan speed 10 μm/s). In a "normal" ferroelectric material it would result in the contrast change from bright to dark (or vice a versa) corresponding to the polarization reversal in the nearby tip areas (see, e.g., [19]). The results on STO surface are completely different: signal near the grain boundary was almost unchanged signifying that the polarization is frozen (imprinted) and cannot be switched even under high electric field.

We believe that our results can be interpreted as follows. Surface polarization can, in principle, exist due to the reconstruction of the STO surfaces [10, 20] and could be different for different grains. However, the calculations of the effective piezoelectric effect on different faces of STO have not been yet perfected and it is hardly believed that the surface reconstruction could occur in the sample cooled down in ambient conditions. More natural reason for the appearance of the polarization near STO surface is the flexoelectric effect. It has been shown by Catalan et al. [21] that the strain gradients and associated electric fields due to flexoelectric effect play an important role in the properties of strained $BaTiO_3$ and $(Ba,Sr)TiO_3$ films. More recently, flexoelectric coefficient was directly measured by bending pure STO crystals and the entire matrix of the flexoelectric coefficients was determined [5]. The coefficients had both positive and negative signs in a qualitative agreement with the results of our measurements (Figs. 1 and 2). Surface flexoelectricity in STO ceramics can be evaluated as follows. It is known that the lattice parameters decrease upon approaching to the STO surface as judged by the asymmetric x-ray diffraction [11]. Thus the value of the strain gradient near the ceramic surface can be roughly estimated as $\frac{\partial u}{\partial x} = 2.5 \cdot 10^5$ m$^{-1}$. Flexoelectric polarization based on the measured average piezoelectric coefficient is approximately 0.7 μC/cm$^2$. Thus the average absolute value of the flexoelectric coefficient is of the order of 0.1 nC/m. For the comparison, the value of the flexoelectric coefficients for (001) orientation of STO is +6.1 nC/m [5], i.e., more than an order of magnitude greater. It is obvious that the piezoresponse histogram shown in Fig. 2 is an envelop of the piezoelectric coefficients of different grains determined via flexoelectric polarization $P^f$ in a general form:

$$d_{ijk} = 2\varepsilon_o Q_{imnk} P_m^f \varepsilon_{lj},$$

where $Q_{imnk}$ is the tensor of electrostriction coefficients, $\varepsilon_o$ is the permittivity of vacuum, $\varepsilon_{lj}$ is the tensor of dielectric permittivities, $P_m^f = f_{mnop} \frac{\partial s_{no}}{\partial x_p}$ is the flexoelectric polarization determined by the strain gradient $\frac{\partial s_{no}}{\partial x_p}$ and tensor of flexoelectric coefficients $f_{mnop}$. It should be mentioned that the matrix of the induced piezoelectric coefficients has to be rotated in accordance with the orientation of a particular grain. From this point of view, the measured deformation is not necessarily normal to the sample surface and can yield both OPP and IPP components. It is possible that higher values for negative flexoelectric coefficients result in the shift of the histogram and asymmetry. Smaller values of the apparent flexoelectric coefficient can come from surface mechanism [22] that is not scaled with the dielectric permittivity and can give discrepancies with the single crystal data [5] which is essentially a bulk effect.

It should be noted that the history of flexoelectric measurements in STO and other high permittivity materials dates back to the beginning of 1980s when it has been shown that the major contribution to thermopolarization effect (polarization appearance under the temperature gradient [6, 23]) comes from flexoelectricity that translates inhomogeneous strain resulted from temperature gradient into macroscopic polarization [6]. In this case, the flexoelectric coefficient could be also estimated from the measured thermopolarization coefficient at 130 K [7] and thermal expansion coefficient for $SrTiO_3$. It gave the value of relevant flexoelectric coefficient of ≈0.2 nC/m that is very close to that measured by PFM taking into account the increase of flexoelectricity with decreasing temperature [5].

It is worth noting that the observed contrast cannot be due to a spurious electrostatic effect. The detailed analysis of tip-surface interactions in OPP PFM of ferroelectric surfaces presented in [16, 24] shows that the uniform piezoelectric-like contrast can be attributed either to the potential variation above the surface and corresponding change of the capacitive interaction or the variation in the surface charge density and normal electric field that results in additional Coulomb interaction between the tip and the surface. Electrostatic effects in OPP PFM measurements diminish with increasing stiffness of the cantilever. However, in our case the contrast did not vary with the cantilever stiffness. This and the existence of clear IPP contrast points to the piezoelectric nature of the signal. Due to the same fact, the appearance of flexoelectric effect awing to high inhomogeneous stress/strain exerted by the PFM tip itself is also ruled out as the piezoresponse signal did not depend not only on the cantilever stiffness but also on the setpoint (e.g., on the dc force applied to the surface).

The increase of the piezoresponse (mostly IPP component, Fig. 2b) near the grain boundaries is the natural result of the redistribution of the mobile defects such as oxygen vacancies and manifests itself in a range of phenomena such as the appearance of forbidden Raman and IR modes [11, 13, 15] and transport across grain boundary [25]. The measured piezoelectric coefficient is somewhat higher than the surface one (1-2 $\mu C/cm^2$) and depends on the tilt between crystallographic axes of the grains [25]. It should be noted that during sintering inhomogeneous localized point defects, e.g., oxygen vacancies may drift to the grain boundaries [26] and result in the strain gradients similar to those observed near the surface.

In order to confirm the piezoelectric nature of the obtained images, we measured IPP contrast at two different orientations of the sample (Fig. 3). Sample rotation by 180° around the axis normal to the surface does not change the topography and electrostatic cross-talk, however, they should change the in-plane piezoresponse reversing it contrast. Observable changes on the IPP image are compatible with piezoelectric nature of the effect in the STO ceramics. This rules out the possible contribution of electrostatic signal to the observed signal.

In conclusion, piezoelectric contrast was observed on the polished surfaces of pure $SrTiO_3$ ceramics. The contrast was attributed to the surface piezoelectricity in $SrTiO_3$ with the polar axis having arbitrary direction with the surface normal. Flexoelectric effect (strain gradient-induced polarization) due to surface relaxation was invoked to explain the observed phenomena. The estimated average flexoelectric coefficient was approximately one order of magnitude smaller as compared to those recently measured in STO single crystals. Notable increase in the in-plane piezoelectric signal near the grain boundaries (not related to topography cross-talk) was observed and explained by the dipole moments due to diffusion of localized defects during sintering.

**Acknowledgments:**
The authors wish to thank Portuguese Foundation for Science and Technology (project PTDC/FIS/81442/2006) and NoE FAME (NMP3-CT-2004-500159) for the financial support. The work was also supported by the Grant Agency of the Czech Rep. (project 202/06/P219) and Czech Academy of Sciences (project AVOZ 10100520). Collaboration with Scientec

8**Figure captions:**

**Figure 1.** AFM images of STO ceramic sample: (a) - topography, (b) - OPP PFM, (c) - IPP PFM. $U_{ac}$ = 30 V, f = 50 kHz.

**Figure 2.** (a) Histogram of the OPP PFM image of STO ceramics, (b) – cross-section of IPP signal along the line AB (Fig.1c). Dotted line represents ratio | IPP / OPP |.

**Figure 3.** PFM images of STO ceramic sample. (a) - IPP signal, (b) – same IPP signal after rotation the sample at 180º.

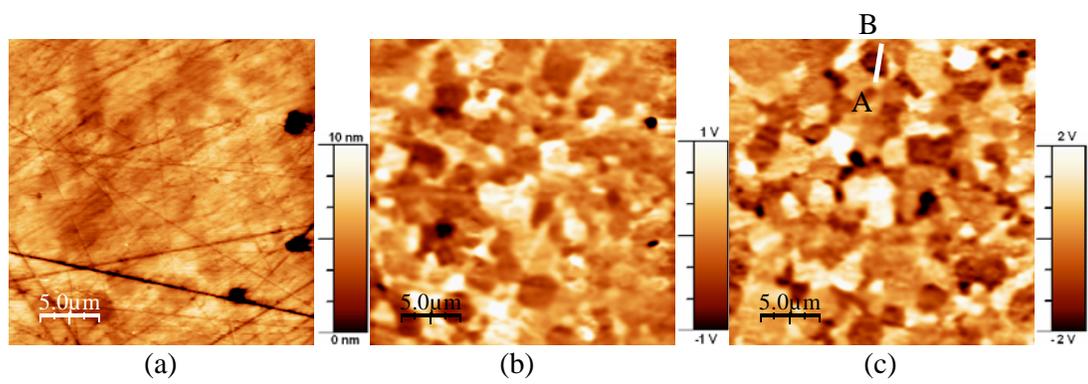

Figure 1.

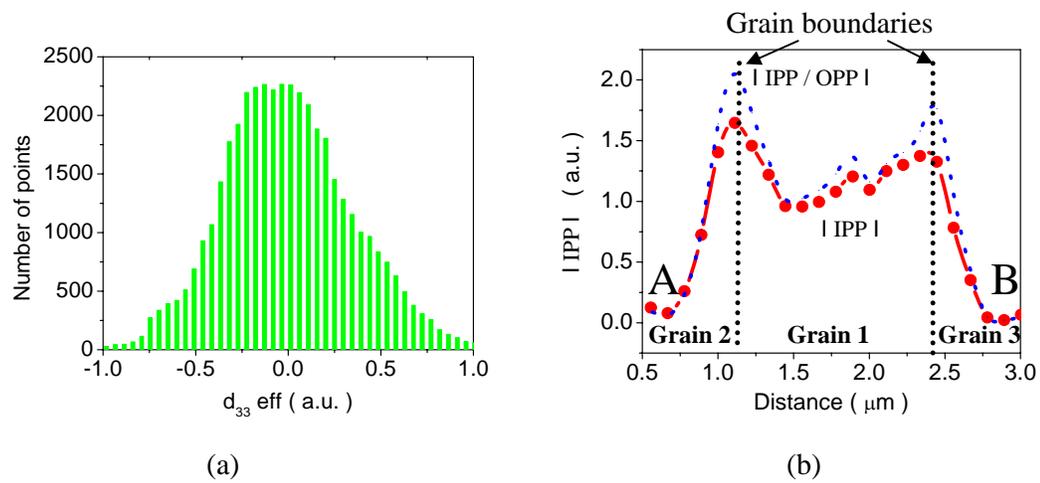

(a)

(b)

Figure 2.

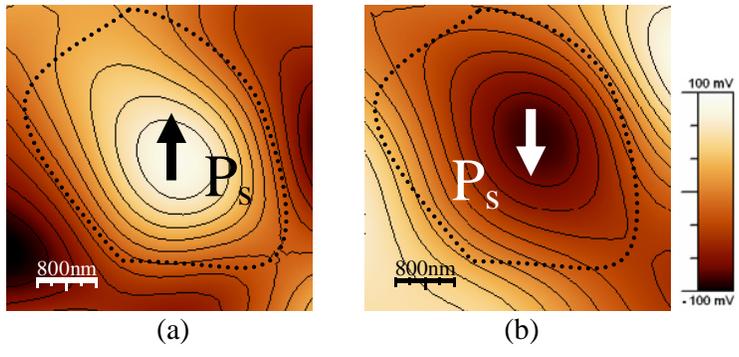

(a)  (b)

Figure 3.